\documentclass[pre,preprint,superscriptaddress]{revtex4}

\usepackage[pdftex]{graphicx}
\usepackage{amsmath}
\usepackage[normalem]{ulem}

\begin{document}

\title{Negative-feedback self-regulation contributes to robust and high-fidelity transmembrane signal transduction}

\author{M. \'Angeles Serrano}
\affiliation{Departament de F\'isica Fonamental, Universitat de Barcelona, Mart\'i i Franqu\`es 1, 08028, Barcelona, Spain}
\author{Manuel Jurado}
\author{Ramon Reigada}
\affiliation{Departament de Qu\'imica F\'isica, Universitat de Barcelona, Mart\'i i Franqu\`es 1, 08028, Barcelona, Spain}

\begin{abstract}
We present a minimal motif model for transmembrane cell signaling. The model assumes signaling events taking place in
spatially distributed nanoclusters regulated by a birth/death dynamics. The combination of these spatio-temporal aspects
can be modulated to provide a robust and high-fidelity response behavior without invoking 
sophisticated modeling of the signaling process as a sequence of cascade reactions and fine-tuned parameters. 
Our results show that the fact that the distributed signaling events take place in nanoclusters with a finite lifetime regulated by local
production is sufficient to obtain a robust and high-fidelity response.
\end{abstract}

\keywords{transmembrane signal transduction, signaling nanoclusters,
high-fidelity transduction, MAPK pathway}


\maketitle

\vspace{-0.5cm}
\section{Introduction}\
Transmembrane cellular signaling pathways are responsible for linking external stimuli and internal cellular actions.
Typically, signaling molecules activate specific receptor proteins on the cell membrane triggering a cascade of interactions
that diffuse messengers inside the cell regulating its activity, growth and development. Signaling pathways not only sense
information but integrate and process it into fast (seconds to minutes) and robust high-fidelity responses ~\cite{bhalla}.
The importance of certain ubiquitous motifs in signaling pathways --such as activation cascades and feedback loops-- for this
processing is well recognized. How spatio-temporal mechanisms influence signal transduction is much less understood. 

The interplay between cascades and feedback loops can explain intriguing properties such as opposing cell fate decisions depending
on different stimuli provoking different activation amplitudes ~\cite{fell,muller,shin,deRonde,recio}. As a striking example,
the mitogen-activated protein kinase (MAPK) pathway governs crucial cellular processes like proliferation and differentiation ~\cite{kolch0},
and its dysfunction has been related with cancer ~\cite{dhillon}.
In short, the pathway consists of the sequential activation of three kinases. 
The transduction process is initiated by a growth-factor-induced recruitment of the SOS factor to the plasma membrane that links and activates
a G-protein. The latter recruits a MAPK kinase kinase (MAPKKK) from the cytosol to the plasma membrane,
that double-phosphorylates and activates a MAPK kinase (MAPKK), that in turns, double-phosphorylates and activates a MAPK as
a final signaling messenger. The most known example of this MAPK cascade corresponds to the regulation of ERK (MAPK) which
features Ras as the G-protein, Raf as the MAPKKK and MEK as the MAPKK ~\cite{kolch0}. 
Mathematical models have shown that, in the absence of feedback regulation, the MAPK cascade elicits a steep response to the input signal
if successive protein activations are performed in a distributive manner ~\cite{ferrell}, while feedback regulations modulate the overall
sensitivity of the pathway ~\cite{kholodenko,kolch}.

\begin{figure}
\centerline{\includegraphics[width=8cm]{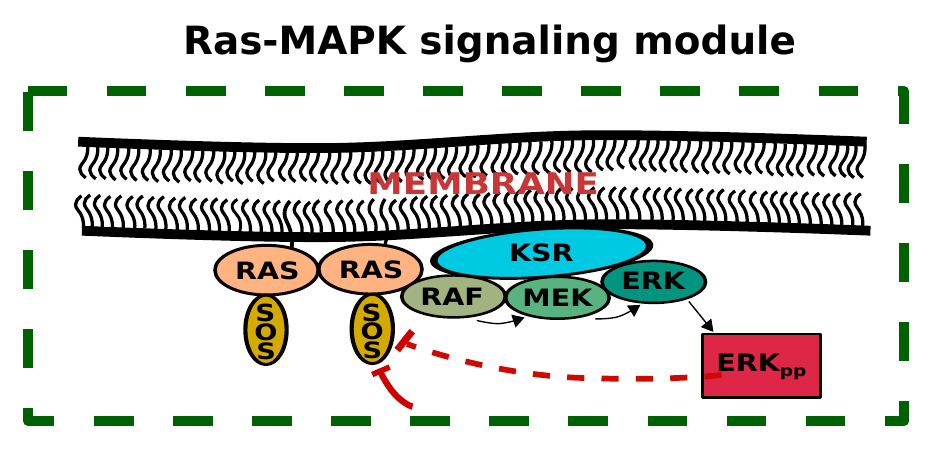}}
\caption{Sketch of the Ras-MAPK signaling platform as described in the literature  ~\cite{kolch0,hancock1}.
Negative-feedback regulations with two different origins have been represented in red: the one in solid line corresponds to a external regulation,
whereas the one in dashed stands for a internal self-regulation.}
\label{figura1}
\end{figure}

Recently, membrane nanoclusters concentrating signaling proteins have been proposed as a new fundamental mechanism to modulate
and increase the efficiency and specificity of the MAPK cascade. The work by Harding and Hancock ~\cite{hancock1}
has revealed that activated receptor proteins aggregate in the membrane forming nanoclusters, that recruit the
downstream factors (Raf, MEK and ERK) from the cytosol and perform signal transduction. These signaling platforms (Fig. \ref{figura1})
display two important spatio-temporal characteristics: they are transient with short lifetimes (typically under $1s$),
and dispersed in the cell membrane occupying a small reaction volume (radii $\approx 10 nm$) ~\cite{hancock1}. When the nanocluster
spatial organization is combined with the corresponding cascades and feedback loops, \emph{in silico} and \emph{in vivo} analyses of MAPK
signaling show that the system is able to perform high-fidelity signal transduction. 
The numerical implementation of the latter proposal by Tian et al  ~\cite{hancock3}, shows that in a range of kinetic parameters,
nanoclusters work as switches responding maximally to very low input signals. Since the generation of signaling platforms is
proportional to the input stimulus, nanocluster ultrasensitivity results in high-fidelity signal transduction (the global response is
proportional to the stimulus). Despite the undeniable relevance of the work of by Tian et al. in Ref. ~\cite{hancock3}, high-fidelity
seems to require a precise selection of the kinetic rates of the reactions involved in the proposed signaling process. Actually, modifications of the
kinetic model parameters may result in poor signal transmission ~\cite{hancock3}. Additionally, a critical analysis of the
contribution of the spatio-temporal nanocluster dynamics on the signal/response behavior is missing.
The intricacy of the signaling cascade and high number of tuned-for-ultrasensitivity parameters in that \emph{in silico} approach
hinders to discriminate the role of the spatio-temporal nanocluster dynamics in signaling output. 

In this Letter, we consider a remarkably simple and generic motif model that incorporates
the aggregation of signaling proteins into discrete transient domains, but that simplifies the pathway structure to unveil the importance
of the spatio-temporal dynamics of membrane nanoclusters. We want to emphasize that nanocluster dynamics
may control the general stimulus/response behavior of signaling processes involving
dispersed signaling platforms, regardless of the particular architecture of their signaling circuits. 
Interestingly, we find that complex behaviors attributed to the particular architecture --cascades of distributed activations
and feedback loops-- of signaling pathways can be also achieved by regulating
the spatio-temporal dynamics of nanoclusters encapsulating much simpler signaling motifs.
More specifically, ultrasensitivity of signaling platforms leading to high-fidelity transmission, is found to be affected by their
lifetime. To explore how cells might have protected the fidelity of signal transduction, two scenarios have been compared: a situation where
the nanocluster lifetime is externally regulated, and the case where it is not prefixed but
linked to their local activity (see Fig. \ref{figura1}). In the latter case, we report that self-regulation
induces robust individual switch-like behavior and high fidelity global responses for a wide range of kinetic parameters.
\begin{figure}
\centerline{\includegraphics[width=8cm]{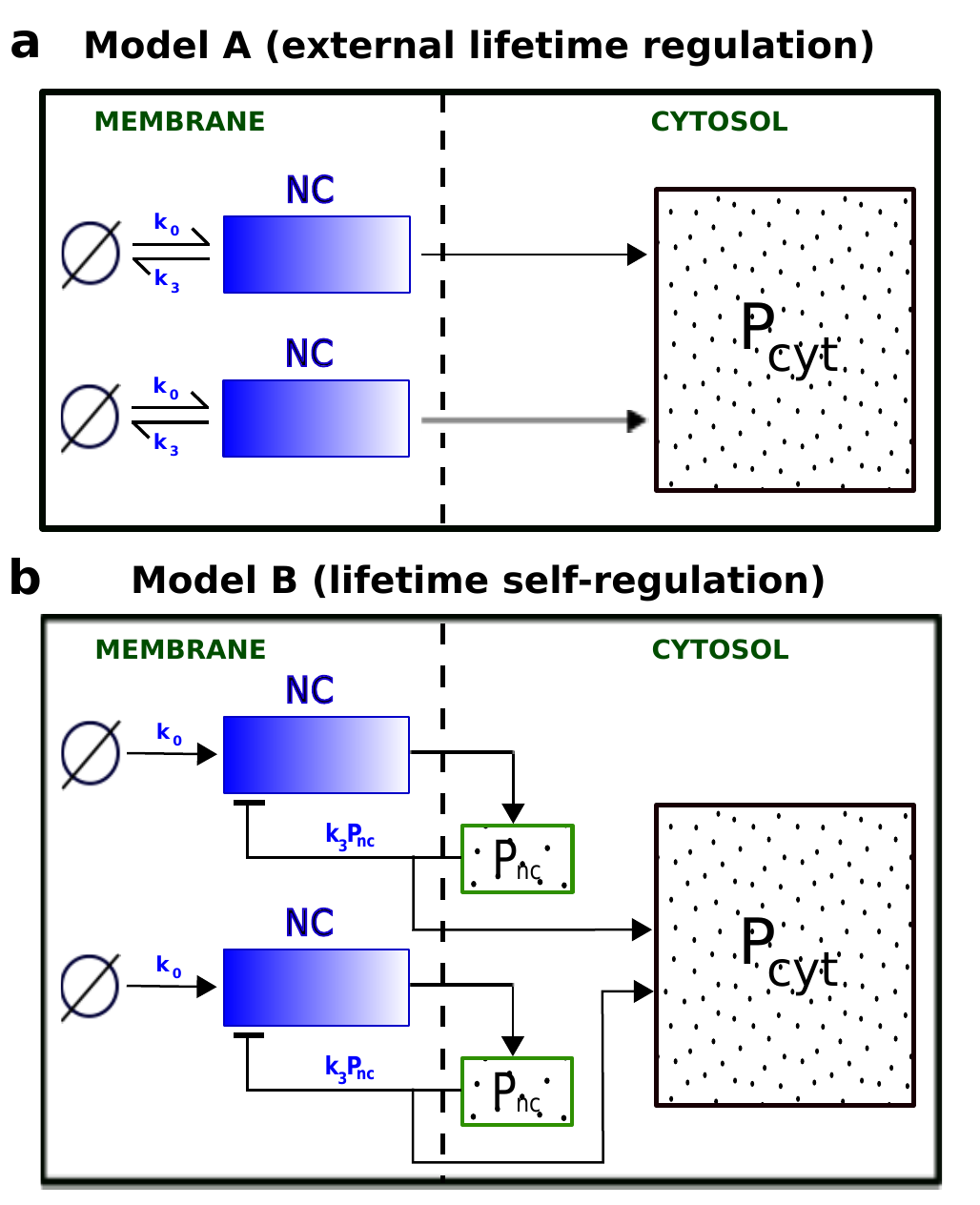}}
\caption{Schematic representation of the two proposed nanocluster ($NC$) disassembling mechanisms.
a) Model A: nanoclusters generate signaling output species that freely diffuse to the cytosol ($P_{cyt}$). Their life-time
is regulated by a frequency determined externally ($k_3$). 
b) Model B: signaling output molecules transiently reside close to the nanocluster ($P_{nc}$) and self-regulate its life-time ($k_3 P_{nc}$).
In both panels only two nanoclusters are represented, but the number in our simulations is on order of hundreds.}
\label{figura2}
\end{figure}

\vspace{-0.2cm}
\section{The model}\
The description of complex signaling circuits as those for the MAPK pathway is often performed by the combination of transformation reactions
modeled as simple enzymatic processes. Therefore, the simplest way to describe the complex features of nanoclusters signal transduction
is to model each nanocluster as a minimal signaling motif based on the standard Michaelis-Menten formulation, 
\begin{equation}
NC+S{\underset{k_{-1}}{\overset{k_{1}}{\rightleftharpoons}}}NC-S\stackrel{k_2}{\rightarrow}P+E \,
\label{eq1}
\end{equation}
where $NC$, $S$, $NC-S$ and $P$ stand for the enzyme, substrate, intermediate and product species, respectively, and  $k_1$, $k_{-1}$ and $k_2$ are
reaction rate constants. The role of the enzyme is assigned here to the nanocluster platform $NC$ where the activity of the signaling
motif takes place. Typically, nanocluster platforms are transient structures assembled by anchored proteins in the inner leaflet of plasma membrane.
These can be proteins like GTPase Ras, that become activated by mediation of the cytoplasmatic protein SOS as a catalyst when
the external stimulus (f.i., growth factor) binds to membrane receptors. In the context of MAPK signaling the first
reversible reaction in Eq. (\ref{eq1}) corresponds to the recruiting of Raf protein to immobile Ras nanoclusters and its subsequent activation,
which starts the MAPK cascade, whereas the second (catalytic) step comprehends successive phosphorylation and activation of MEK and ERK kinases.

In our model, signaling nanoclusters are assumed to follow a dynamics that controls their number and lifetime.
We first consider a birth/death mechanism that does not depend on the spatial distribution or the functioning of nanoclusters,
but represents some extrinsic regulation (model A, Fig. \ref{figura2}a). 
Nanoclusters are dynamically formed in the cell membrane at rate $k_0$, whereas, independently, a fixed constant rate $k_3$ determines
the frequency of nanocluster disassembling. During their lifetime, signaling nanoclusters can generate product molecules
according to the reaction motif in Eq. (\ref{eq1}). 
The external stimulus is represented by parameter $\alpha$, that is set to unity at maximal stimulus.
The role of $\alpha$ is two-fold. First, in vivo experiments reveal that nanocluster generation is proportional to input growth
factor concentration ~\cite{hancock3}, so we consider the frequency of nanocluster formation to be proportional to the stimulus,
$k_0= \alpha k_0^{(m)}$, where $k_0^{(m)}$ corresponds to nanocluster generation rate at maximal stimulus. Second, kinase phosphorylation is known
to happen in two separated encounters, both promoted by stimulus, instead of occurring sequentially in a single encounter ~\cite{ferrell}.
Such a distributive mechanism is accounted in our signaling motif by setting $k_2= \alpha k_2^{(m)}$, with $k_2^{(m)}$ being
the catalytic rate at maximal stimulus (see also Refs. ~\cite{ferrell,hancock3}).

In the numerical simulations, we follow a stochastic approach similar to the Gillespie algorithm ~\cite{gillespie,gillespie1}.
All events corresponding to nanocluster birth and death,
and to internal molecular transformations, are treated as stochastic events because of the small numbers of proteins
involved and the limited lifetime of signaling platforms. In particular, we consider Poissonian processes with a frequency
determined by the corresponding rates, that can be found in literature. For the MAPK pathway, cytosolic concentration of
Raf is about $10^{-7} M$ ~\cite{ferrell}, whereas Raf activation is of order $\approx 10^6-10^8 M^{-1}s^{-1}$ ~\cite{muller,kolch},
which leads to a forward reaction rate $k_1 \approx 0.1-10 s^{-1}$. The dissociation reaction is slower,
$k_{-1}\approx10^{-2} s^{-1}$~\cite{muller,kolch}. Catalytic constant rates corresponding to kinase phosphorylation processes
are much larger, $k_{2}\approx10-100 s^{-1}$ ~\cite{muller,kolch,hancock3}. Finally, death rate $k_3=2s^{-1}$ is tuned
to adjust the estimated $0.5 s$ average lifetime of Ras nanoclusters ~\cite{hancock3}.
Activation frequency at maximal stimulus, $k_0^{(m)}$, is arbitrarily fixed to $1000 s^{-1}$, so that the average number
of simulated nanoclusters is of the order of a few hundreds.
This number is well below the typical maximum number of Ras nanocluster in a cell ($\approx 50000$) ~\cite{hancock3}
but assures a sufficient statistical ensemble for our stochastic simulations in a reasonable computational time.
\begin{figure}
\centerline{\includegraphics[width=8cm]{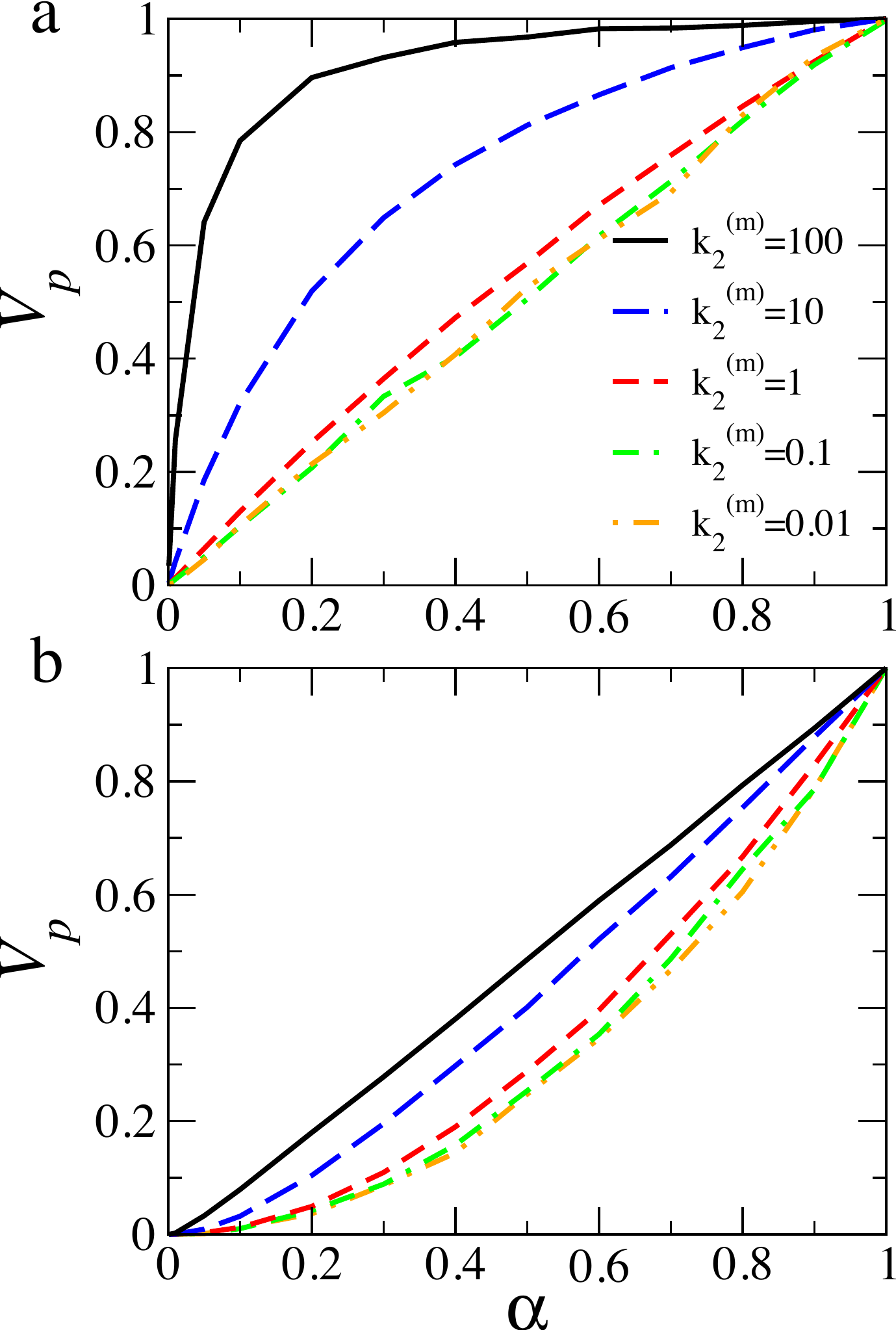}}
\caption{Global response $V_P$ as a function of stimulus $\alpha$ in
model A (external lifetime regulation), for $k_1=1s^{-1}$. a) Number of nanoclusters fixed and independent of the input stimulus
($k_0= k_0^{(m)}$). b) Number of nanoclusters regulated by stimulus ($k_0= \alpha k_0^{(m)}$).}
\label{figura3}
\end{figure}

\vspace{-0.5cm}
\section{Results}\
The stimulus/response behavior of the model is evaluated by representing the velocity $V_P$ of $P$ formation normalized
with respect to the maximal production (for $\alpha =1$) as a function of input stimulus $\alpha$. 
For model A, individual signaling nanoclusters can generate both graded and switch-like outputs (see. Fig. \ref{figura3}),
depending on the relative values of the reaction rates
and whenever nanoclusters do not die too fast, $k_1\lesssim k_3$. If the limiting reaction is the formation of product molecules from
the intermediate complex $NC-S$, $k_1 > k_2$, nanoclusters generate a graded signal since the impact of the distributive mechanism
in the formation of product $P$ through $k_2= \alpha k_2^{(m)}$ becomes relevant.
In contrast, for $k_1 < k_2$, the limiting reaction is the formation of $NC-S$. In this situation, once the intermediate complex
is formed, product molecules are generated very fast, quite independently of the contribution of the stimulus in the catalytic step.
Then, a nanocluster responds maximally to very low inputs (ultrasensitivity)
and its output is switch-like, presenting a steeper response for greater differences between $k_1$ and $k_2$.
These behaviors are summarized in Fig. \ref{figura3}a, that reports results from simulations run with a fixed number of operating nanoclusters
independent of the input stimulus; namely, for a nanocluster generation rate $k_0$ fixed to $k_0^{(m)}$ in order to average a large
ensemble of signaling platforms. The stimulus/response curves are plotted for different values of $k_2^{(m)}$, given $k_1=1s^{-1}$.
\begin{figure}
\centerline{\includegraphics[width=8cm]{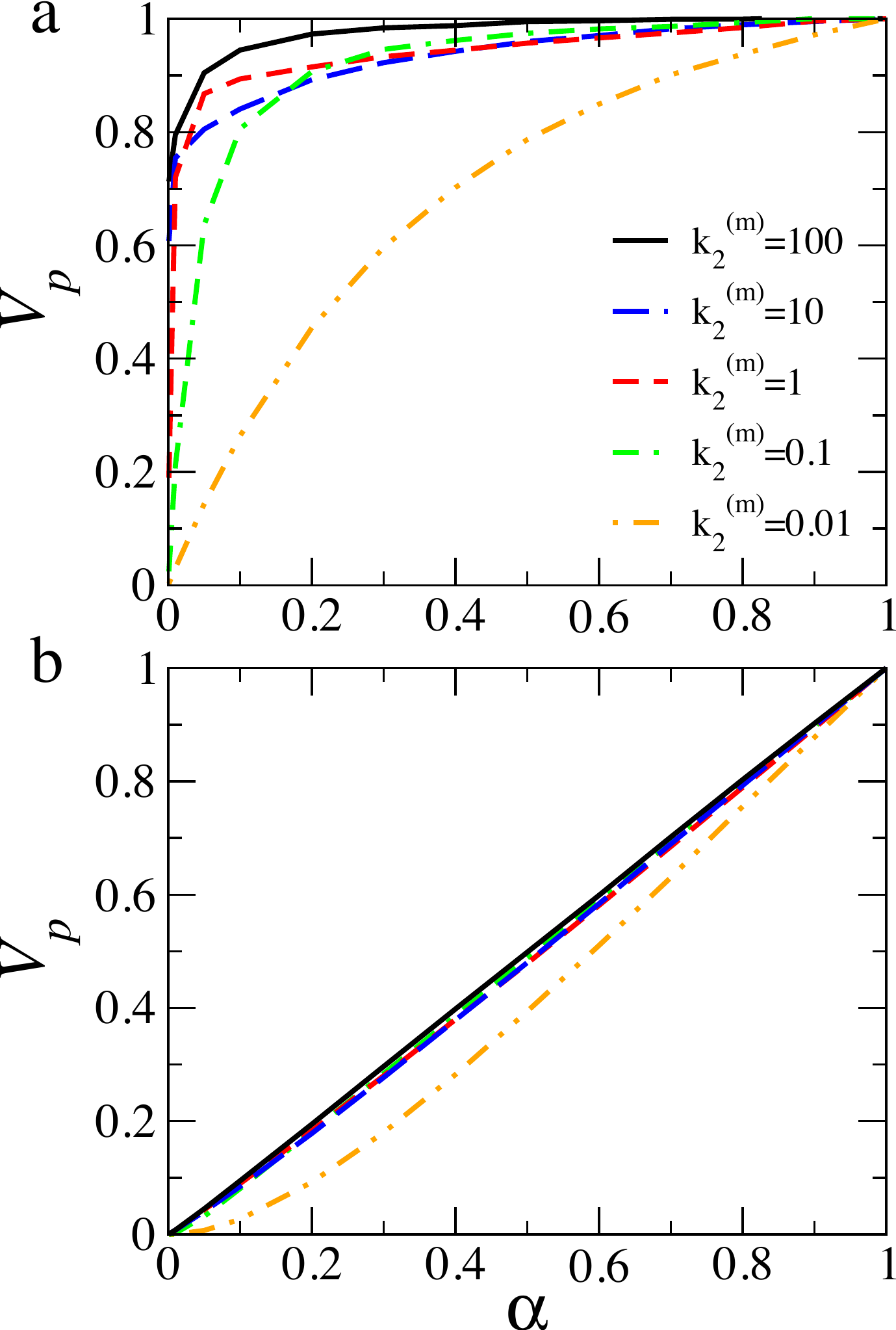}}
\caption{Global response $V_P$ as a function of stimulus $\alpha$ in
model B (lifetime self-regulation), for $k_1=1s^{-1}$. a) Number of nanoclusters fixed and independent of the input stimulus
($k_0= k_0^{(m)}$).
b) Number of nanocluster regulated by stimulus ($k_0= \alpha k_0^{(m)}$).}
\label{figura4}
\end{figure}
The global response of the system corresponds to the integration of local signaling events taking place in different activated nanoclusters.
In the cell membrane, the number of nanoclusters is proportional to stimulus concentration, $k_0= \alpha k_0^{(m)}$ ~\cite{hancock3}.
Figure \ref{figura3}b shows that this global response is nonlinear when the nanoclusters function in the graded regime
(low values of $k_2^{(m)}$), while the system generates a graded nearly linear product output in the regime of nanoclusters
working as nanoswitches (large values of $k_2^{(m)}$), meaning that in this case the signaling response is directly proportional
to input stimulus. The fact that nanoclusters respond maximally even at low stimulus results in a global response only dependent
on the number of nanoclusters, proportional to stimulus. This implies that the system of ultrasensitive nanoswitches
achieves high-fidelity signal transduction by performing an analogue-digital-analogue transmission ~\cite{hancock3}, an obvious
advantage that could be one of the reasons for the active compartmentation of cell signaling processes.
We note that our results are obtained for the simplest internal structure of the signaling pathway given in Eq. (\ref{eq1}),
without the recursion to intricate reactions profuse in biological details.

As an alternative to model A, we propose a signaling mechanism where the nanocluster lifetime is self-regulated
via the coupling of nanocluster disassembly to its local activity, without the intervention of external factors (model B, Fig. \ref{figura2}b).
We assume that product molecules remain transiently accumulated in a local pool that acts as an internal clock exerting a negative
impact in the lifetime of the nanocluster.
The more it has produced, the higher its chances to die, as it happens for instance in Ras-MAPK signaling: production of ERK promotes
the phosphorylation of the SOS factor, which inhibits the activity of Ras aggregates ~\cite{kolch0,shin}.
The concept of local pool is acceptable for short nanocluster lifetimes of fractions of second. Then, the restricted diffusive motion of product species does not allow them to travel far from the signaling membrane complex.
We model the death rate of a nanocluster as $k_3 P_{nc,i}(t)$, where $k_3$ is a constant rate and $P_{nc,i}(t)$ is the number of signaling
product molecules in the local pool generated by the nanocluster $i$ up to time $t$. The latter expression is the simplest dependence
for a negative upstream inhibition due to the final product protein and corresponds to the opposite limit to model A. Whereas model A considers
that the dynamics of signaling platforms is completely regulated externally, in model B this regulation is absolutely modulated by their local activity.

To investigate the impact of local activity regulating the death of nanoclusters, we study the response of model B
for the same parameter values used in the simulations of model A in Figs. \ref{figura3}a and \ref{figura3}b.
Lifetime self-regulation enhances nanocluster sensitivity, maintaining the switch-like response for a wider range of values of $k_2^{(m)}$,
see Fig. \ref{figura4}a for a fixed number of signaling platforms independent of stimulus ($k_0= k_0^{(m)}$).
The explanation is related to the fact that in model B nanoclusters can be active for longer periods when the input signal is low. In model A,
most nanoclusters are disassembled before producing any $P$ molecule when working under a low stimulus. In these cases, and in particular when $k_2^m$
is small, the effect of the input stimulus in the production of $P$ becomes critical.
Instead, in model B, the death probability of a nanocluster is zero until it has produced at least one P molecule, so that the second step in Eq. (\ref{eq1})
is no longer critical even at low values of $k_2^{(m)}$. Notice that the lifetimes of nanoclusters are longer in model B
at low input stimulus, they approach the lifetime fixed in model A ($k_3^{-1}$) at moderate stimulus, and may become even shorter at maximal stimulus.
As a consequence of the gain in nanocluster sensitivity, the global response fidelity is also enhanced in model B, as
it is shown in Fig. \ref{figura4}b for simulations using $k_0= \alpha k_0^{(m)}$.

\begin{figure}
\centerline{\includegraphics[width=8.5cm]{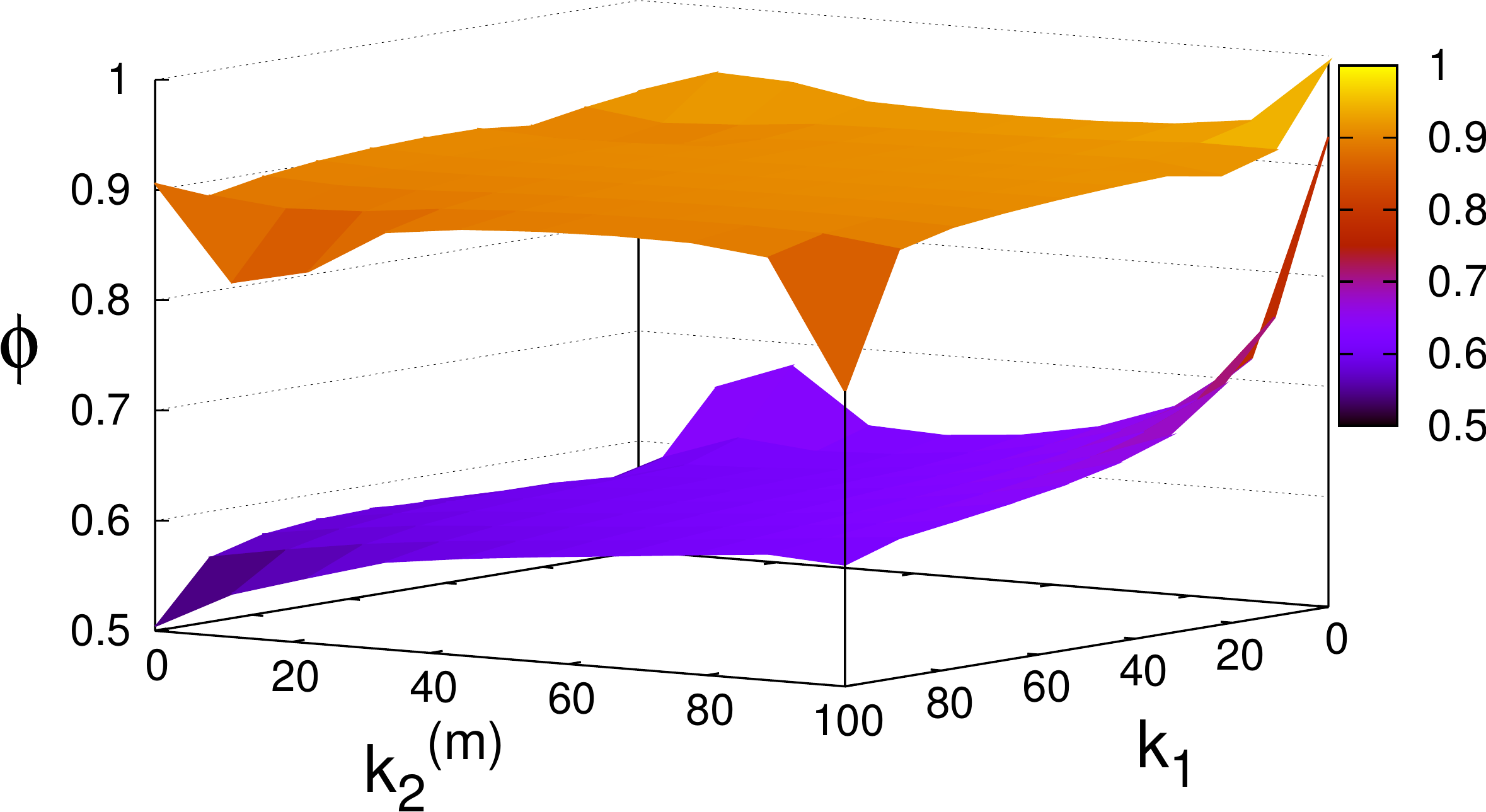}}
\caption{Signal transmission fidelity $\phi$ computed at different values of $k_1$ and
$k_2^{(m)}$ for models A (bottom surface) and B (top surface). Simulations using $k_0= \alpha k_0^{(m)}$.}
\label{figura5}
\end{figure}

We have also performed a systematic study of signal transmission fidelity in both models at different values of the kinetic rates $k_1$ and $k_2^{(m)}$.
Fidelity of signal transduction is quantified here by parameter
$\phi$, computed as the integral of the global response rescaled by the number of nanoclusters, $\frac{V_P}{\alpha}(\alpha)$, for the whole range of input stimulus. High-fidelity
transduction is achieved when the system response, $V_P$, correlates to stimulus, $\alpha$, so when $\phi \rightarrow 1$.
This requires nanoclusters to work in the range where they behave as ultrasensitive nanoswitches.
Figure \ref{figura5} presents the values of $\phi$ in the ($k_1$,$k_2^{(m)}$) parameter space spanning two orders of magnitude for each kinetic rate.  Notice that high-fidelity signal transmission is a robust feature for the signaling mechanism involving a self-regulated nanocluster dynamics (model B),
whereas for the scheme based on a fixed nanocluster lifetime (model A) such virtue requires a fine tuning of the reaction kinetic parameters.

\vspace{-0.5cm}
\section{Conclusions}\
Two important conclusions can be derived from our simulations. First, the sensitivity of signaling platforms
is found to be modulated by their lifetime, so it is not exclusively determined by the particular architecture of the signaling pathway as suggested so far.
Second, comparison of two extreme models for nanocluster disassembly reveals the importance of the physical origin of nanocluster lifetime regulation. As two extreme possibilities, we propose model A where nanoclusters lifetime is externally regulated by a fixed frequency, and model B
where individual nanocluster activity fully determines its duration. Most likely, biological cell signaling may be regulated by a mixture
of these two extreme situations.  Importantly, we have shown that any contribution to nanocluster lifetime regulated by local
production promotes robust individual ultrasensitivity outputs and, consequently, high-fidelity global responses for a wider range of reaction
kinetic rates as compared to model A. Therefore, it could be conjectured that nanocluster lifetime self-regulation protects the signaling response from
variability in the particular architecture of the signaling structure and in the rates of involved reactions.
Many modeling approaches have been attempted to describe cell signaling processes but the complexity of the signaling structure and the number of
kinetic parameters hide the particular role of nanocluster dynamics. In this Letter,
we have proposed a simple and generic signaling motif model that captures essential features of signal transduction, such as ultrasensitiveness and fidelity, and could apply to different types of spatial signaling domains following a temporal birth/death dynamics.
Nevertheless, the observed complexity of signaling pathways may entail some biological advantages, such as enabling plasticity to modulate
different response amplitudes triggering opposing cell fate decisions within a single cell. In the context of our framework for nanocluster
cell signaling, we plan to tackle these questions and others, such as the processing of time dependent stimulus, in future research.

\vspace{-0.6cm}
\section*{Acknowledgments}
\vspace{-0.4cm}
This work was supported by Generalitat de Catalunya grant Ref. 2009SGR1055; the Ram\'on y Cajal program of MICINN; MICINN Project BFU2010-21847-C02-02.

\end{document}